\documentclass[10pt, a4paper]{article}

\usepackage{lrec-coling2024} % this is the new style

\usepackage{tcolorbox}
\tcbuselibrary{raster,skins,breakable}
\tcbuselibrary{xparse} %preamble
\DeclareTColorBox{brekableitembox}{ o m O{.5} O{} }% 
  {empty, left=2mm, right=2mm, top=-2mm, bottom=0.2mm, attach boxed title to top left={xshift=1.2em},
  boxed title style={empty,left=-2mm,right=-2mm}, colframe=black, coltitle=black, coltext=black, breakable,  
  underlay unbroken={\draw[black,line width=#3pt]
    (title.east) -- (title.east-|frame.east) -- (frame.south east) -- (frame.south west) -- (title.west-|frame.west) -- (title.west); },
  underlay first={\draw[black,line width=#3pt](title.east) -- (title.east-|frame.east) -- (frame.south east) ;
    \draw[black,line width=#3pt] (frame.south west) -- (title.west-|frame.west) -- (title.west); },
  underlay middle={\draw[black,line width=#3pt](frame.north east) -- (frame.south east) ;
    \draw[black,line width=#3pt](frame.south west) -- (frame.north west) ;},
  underlay last={\draw[black,line width=#3pt](frame.north east) -- (frame.south east) -- (frame.south west) -- (frame.north west) ;}, IfValueTF={#1}{title=~~#2~~〈#1〉}{title=~~#2~~},#4}

\title{Construction of a Japanese Financial Benchmark\\for Large Language Models}

\name{Masanori Hirano} 

\address{Preferred Networks, Inc. \\
         Tokyo, Japan \\
         research@mhirano.jp}

\abstract{
With the recent development of large language models (LLMs), models that focus on certain domains and languages have been discussed for their necessity.
There is also a growing need for benchmarks to evaluate the performance of current LLMs in each domain.
Therefore, in this study, we constructed a benchmark comprising multiple tasks specific to the Japanese and financial domains and performed benchmark measurements on some models.
Consequently , we confirmed that GPT-4 is currently outstanding, and that the constructed benchmarks function effectively.
According to our analysis, our benchmark can differentiate benchmark scores among models in all performance ranges by combining tasks with different difficulties.
\\ \newline  }

\begin{document}

\maketitleabstract

\section{Introduction}
Recently, Large Language Models (LLMs) have demonstrated excellent performance.
In particular, the latest models, such as ChatGPT\cite{chatgpt} and GPT-4\cite{GPT4}, exhibit high performance and significant generalization abilities.
The basis of these models begins with the transformer \cite{Vaswani2017} and BERT\cite{Devlin2018}, and GPT series \cite{GPT-1,GPT-2,GPT-3} were developed using the transformer.
Other LLMs have also been proposed, such as Bard\cite{bard}, LLaMA\cite{touvron2023llama,Touvron2023}, Dolly\cite{dolly}, BLOOM\cite{scao2022bloom}, Vicuna\cite{vicuna}, PaLM\cite{Chowdhery2022,Anil2023}, and Gemini \cite{gemini}.

The major difference between the latest LLMs and previous language models, such as BERT, is that one model can answer questions in multiple languages and domains and respond to questions by following the instructions.
Previously, BERT was trained separately in different languages and domains \cite{Suzuku2023-ipm}.
However, the latest LLMs, such as GPT4, can freely process multiple languages.
Moreover, whereas BERT can only fill in incomplete sentences, the latest LLMs can answer questions in the same manner as humans.

Because of these improvements, the evaluation tasks should be reconstructed.
The latest LLM performances far exceed those of previous language models regarding the variety and accuracy of questions they can answer.
Therefore, a greater variety of questions is necessary to evaluate LLMs more accurately.
Thus, evaluation tasks are important for developing high-performance LLMs.

Currently, some evaluation tasks for LLMs have already been prepared, but are insufficient as concerns domain-specified tasks and those for languages other than English.
For instance, a language model evaluation harness (lm\_eval) \cite{eval-harness} was proposed for LLM evaluation using several English tasks.
Moreover, several domain-specified tasks have been evaluated using GPT-4\cite{GPT4}.
\citet{Eulerich2023} evaluated it using certified public accountant (CPA) tests, \citet{Nori2023} tested it in the medical domain, and its applications to legal services were also tested \cite{Iu2023,Choi2023}.
However, only a small number of domain-specified tasks have been tested, and the response of LLMs to other tasks is still being investigated comprehensively.

This study focuses on evaluations of the Japanese financial domain.
Financial services are relatively large as concerns money spendings.
Moreover, according to World Bank data\footnote{\url{https://data.worldbank.org/indicator/CM.MKT.LCAP.CD}}, Japan has the third-largest listed capital market in the world as of 2020.
Therefore, the usability of LLMs in Japanese and financial domains is a crucial issue.

Several studies have been conducted on Japanese LLMs.
Various models such as CyberAgent's CALM series, Rinna's model, stabilityai's stablelm series, Elyza's model, Preferred Networks' Plamo\texttrademark, and LLM-jp-13B have been proposed.
However, few models have been published in academic research papers, and their performances have not been thoroughly evaluated.
Other studies have tuned existing English-based models to specialize in Japanese-language use\cite{Hirano2023-nbis,sukeda2023jmedlora,suzuki2023base}.
As for the Japanese task evaluation for LLMs, several benchmarks are available, including the jlm\_eval\cite{jlm_eval}, llm-jp-eval\cite{llm-jp-eval}, and Rakuda benchmarks\footnote{\url{https://yuzuai.jp/benchmark}}.

However, no benchmarks or LLMs are specified for both Japanese and financial domain.

Thus, this study proposes a new benchmark for the Japanese financial domain and evaluates several models specified for Japanese.
The benchmark and performance results of the models are publicly available at \url{https://github.com/pfnet-research/japanese-lm-fin-harness}.

\section{Related Works}
Studies on specialized language models in finance and Japanese have been conducted for a long time.
The classic vector embedding technique used in language processing is word2vec \cite{Mikolov2013a}.
Word2vec has also been used in the financial domain \citet{Hirano2019-information}.
After word2vec, ELMo \cite{peters2018elmo}, which uses a bidirectional long short-term memory (LSTM) \cite{schuster1997bilstm} to pre-train a distributed representation, appeared, along with transformer \cite{Vaswani2017}, which is a good alternative to LSTM in time-series processing, and transformer-based BERT \cite{Devlin2018}.

In contrast, the methodologies to fit language models to specific languages or domains are also pursued.
For instance, \citet{howard2018ulmfit} proposed universal language model fine-tuning.
Following this study, some domain- or language-specific language models were developed, such as SciBERT \cite{beltagy2019scibert}, MedBERT \cite{rasmy2021med}, Japanese BERT\footnote{\url{https://huggingface.co/tohoku-nlp/bert-base-japanese}}, and Japanese financial BERT \cite{Suzuki2022-sigfin28}.
Moreover, the methodologies and effects of domain-specified fine-tuning were discussed in\cite{gururangan2020don,Suzuku2023-ipm}.

In the era of LLMs, although several transformer-based language models have been proposed, as described in the Introduction section, several unknown mechanisms of LLMs exist and numerous trials have been performed.

Several proposed LLMs that focus specifically on finance exist.
For instance, BloombergGPT\cite{Wu2023} is a private LLM focused on finance.
In addition, publicly available models, such as FinLLAMA\cite{Fin-LLAMA}, which is a tuned version of LLaMA\cite{touvron2023llama}, FinGPT\cite{yang2023fingpt}, and Instruct-FinGPT\cite{zhang2023instruct}, exist.

Japanese-focused LLMs and benchmarks have also been developed, as mentioned in the Introduction section.

However, currently, no LLMs and benchmarks focused on the Japanese financial domain exist.
Therefore, in this study, we construct a benchmark.

\section{Japanese Financial Benchmark Dataset}

We construct a new Japanese financial benchmark for LLMs, comprising the following five benchmark tasks:
\begin{itemize}
    \item chabsa: Sentiment analysis task in the financial field.
    \item cma\_basics: Fundamental knowledge questions in securities analysis.
    \item cpa\_audit: Tasks on auditing in the Japanese Certified Public Accountant (CPA) exam.
    \item fp2: Multiple choice questions for 2nd grade Japanese financial planner exam.
    \item security\_sales\_1: Practice exam for the 1st grade Japanese securities broker representative test.
\end{itemize}
For chabsa and cpa\_audit, we constructed a dataset using corpora from previous studies.
We constructed the remaining tasks by crawling and cleansing the documents available on the Internet.
In the following section, we describe these tasks in detail.
For each task, an example prompt is shown below, but this is only for illustrative purposes.
Several other types of prompts were also prepared, and those prompts were originally written in Japanese.
For details of the prompts, please refer to the aforementioned public repository.

\subsection{chabsa: Sentiment Analysis Task in the Financial Field}
chabsa \citeplanguageresource{chabsa} is a task to determine the sentiments of specific words with respect to sentences contained in securities reports.
In Japan, listed companies publish securities reports annually.
These data are available from \url{https://github.com/chakki-works/chABSA-dataset}.
Three types of sentiments exist: positive, negative, and neutral.
However, the number of neutral words is extremely small, which may hinder a stable performance evaluation.
Therefore, we decided to treat it as a binary classification task, that is, positive or negative classification.
This implies that data tagged as "neutral" will be regarded as incorrect regardless of whether the output is positive or negative.
Because all the questions were two-choice questions, a random response would yield approximately 50\% correct answers.
For the final evaluation values, we employed the macro-f1 value.
In this dataset, 4334 positive, 3131 negative, and 258 neutral responses were observed.
Therefore, the random response yields an f1 value of 49.15 points.

\begin{brekableitembox}{An example of chabsa}
    Please indicate the sentiment of the targeted word in the following sentences, whether positive or negative.\\
    \\
    Sentence: The Japanese economy continued to gradually recover during the fiscal year ending March 31, 2012.\\
    Target Word: Japanese economy\\
    Answer: positive
\end{brekableitembox}

\subsection{cma\_basics: Fundamental Knowledge Questions in Securities Analysis}
cma\_basics questions basic knowledge in securities analysis.
It was created by crawling and cleansing sample questions from the securities analyst examination.
Therefore, it differs from the first and second rounds of the Japanese securities analyst examination administered by the Securities Analysts Association of Japan.
However, it has the same characteristics as the first-round test, including a multiple-choice format.
In addition, questions containing figures were deleted and the tables were translated into a markdown format.
Since all questions had four choices, randomly selecting an answer results in 25.00\% accuracy.
\begin{brekableitembox}{An example of cma\_basics}
    Please answer the letter corresponding to the appropriate choice for the following question.\\
    \\
    Question:\\
    Which of the following statements about the Japanese economy is incorrect? \\
    A: Real GDP (real gross domestic product) is the level of production activity excluding the effects of price fluctuations.\\
    B: Inflation implies a sustained increase in the general price level. \\
    C: Indirect finance is a form of financial intermediation in which banks and other financial intermediaries play a central role in mediating money lending and borrowing.\\
    D: The fiscal policy of the Bank of Japan adjusts the price level through an increase or decrease in money supply.\\
    \\
    Answer:\\
    D
\end{brekableitembox}

\subsection{cpa\_audit: Tasks on Auditing in the Japanese CPA Exam}
cpa\_audit is a collection of short-answer questions on audit theory from the Japanese CPA examination, and data from a previous study \citeplanguageresource{Masuda2023} were used.
It contains 360 questions with six choices and 38 questions with five choices.
Therefore, 16.98\% of the questions could be answered correctly if they are answered randomly.
\begin{brekableitembox}{An example of cpa\_audit}
    Please answer the letter corresponding to the appropriate combination of symbols to answer the following questions:\\
    \\
    Question: \\
    Choose the most appropriate combination of the following statements regarding CPA audits. \\
    (i) In a stock company, the management has a fiduciary responsibility to properly manage and invest the capital contributed by shareholders and provide an accounting report to shareholders regarding the results of this management responsibility. CPA audits of these financial reports contribute to proper management accountability.\\
    (ii) CPA audit not only plays a role in ensuring the reliability of financial information but also supports corporate governance because it encourages the correction of internal control deficiencies and fraudulent acts discovered in the process.\\
    (iii) As listed companies have a significant influence on society, special provisions are placed on CPAs who audit listed companies, such as the prohibition of independent audits, prohibition of certain non-audit attestation services, and restrictions on employment.\\
    (iv) Because a listed company can raise funds widely from general investors, several interested parties arise, and protection against them is necessary. Therefore, establishing a management system for timely and appropriate disclosure of information to stakeholders is necessary. Therefore, CPAs must perform an internal control audit when a company is newly listed.\\
    \\
    Choices:\\
    A: (i) and (ii)\\
    B: (i) and (iii)\\
    C: (i) and (iv)\\
    D: (ii) and (iii)\\
    E: (ii) and (iv)\\
    F: (iii) and (iv)\\
    \\
    Answer:\\
    A
\end{brekableitembox}

\subsection{fp2: Multiple Choice Questions for 2nd Grade Japanese Financial Planner exam}
fp2 is the choice question for a 2nd grade Japanese financial planner exam.
The past questions from the Japan FP Association’s 2nd grade financial planning skills examination from May 2021 to September 2023 were obtained from the official HP\footnote{\url{https://www.jafp.or.jp/exam/mohan/}} and processed.
Questions containing figures were removed, and the tables were translated into a markdown format.
Because all the questions had four choices, a random answer yielded 25.00\% correct answers.
\begin{brekableitembox}{An example of fp2}
    Please select the appropriate answer to the following question using numbers from 1 to 4:\\
    \\
    Question:\\
    Which of the following statements regarding the conduct of financial planners ("FP") toward their clients is most inappropriate as concerns the relevant laws and regulations?\\
    1. Mr. A, an FP who is not qualified as a lawyer, was consulted by a client about adult guardianship and provided a general explanation on the difference between legal and voluntary guardianship.\\
    2. Ms. B, who is not a licensed tax accountant, received a client's consultation regarding the deduction of medical expenses for income tax purposes and explained that the amount of medical expenses paid, which is compensated for by insurance proceeds, is not deductible as a medical expense deduction.\\
    3. Mr. C, an FP who is not a licensed social insurance consultant, received consultation from a client regarding the deferral of receipt of the basic old-age pension and estimated the pension amount in the case of deferral based on the estimated amount of pension receipt in the client's pension benefit report.\\
    4. Mr. D, an FP who is not registered as a financial instruments business operator, concluded an investment advisory contract regarding asset management with the client and recommended the purchase of individual stocks that were expected to rise in value.\\
    \\
    Answer:\\
    4
\end{brekableitembox}

\subsection{security\_sales\_1: Practice Exam for the 1st Grade Japanese Securities Broker Representative Test}
security\_sales\_1 is a practice exam task that corresponds to the first level of the Japanese securities broker representative test.
It was created by crawling and cleansing to obtain practice examinations and sample questions for the 1st-grade Japanese securities broker representative test.
Consequently, some differences in the question structure and difficulty levels from official Japanese securities broker representative tests exist.
It contains 29 questions with four choices and 28 questions with two choices.
Therefore, even if the questions were answered randomly, 37.28\% of correct answers could be obtained.
\begin{brekableitembox}{An example of security\_sales\_1}
    Please answer the letter corresponding to the appropriate choice for the following question.\\
    \\
    Question:\\
    Please answer if the following statement is correct or incorrect: \\
    A securities broker representative is deemed to have the authority to perform all judicial acts on behalf of the financial instrument firm to which they belong with respect to acts prescribed by law, such as the purchase and sale of securities. \\
    \\
    Choices:\\
    A: Correct\\
    B: Wrong\\
    \\
    Answer:\\
    B
\end{brekableitembox}

\begin{table*}[htbp]
    \centering
    \footnotesize
    \tabcolsep 3pt
    \caption{All Benchmark Results. Some low-performance models are omitted. See full results at the repository as previously mentioned}
    \label{tab:benchmark}
    \begin{tabular}{lcccccc}
        \hline
        Model                                                                                                                                               & Ave.  & chabsa & cma\_basics & cpa\_audit & fp2   & security\_sales\_1 \\
        \hline
        openai/gpt-4-32k                                                                                                                                    & 66.27 & 93.16  & 81.58       & 37.44      & 50.74 & 68.42              \\
        openai/gpt-4                                                                                                                                        & 66.07 & 93.20  & 78.95       & 37.69      & 50.32 & 70.18              \\
        openai/gpt-4-turbo                                                                                                                                  & 64.59 & 92.86  & 76.32       & 36.18      & 50.95 & 66.67              \\
        \hyperlink{https://huggingface.co/Qwen/Qwen-72B}{Qwen/Qwen-72B}                                                                                     & 62.18 & 92.36  & 78.95       & 32.91      & 40.00 & 66.67              \\
        \hyperlink{https://huggingface.co/Qwen/Qwen-72B-Chat}{Qwen/Qwen-72B-Chat}                                                                           & 57.89 & 92.52  & 78.95       & 29.90      & 28.42 & 59.65              \\
        \hyperlink{https://huggingface.co/rinna/nekomata-14b}{rinna/nekomata-14b}                                                                           & 56.03 & 89.70  & 63.16       & 25.13      & 42.53 & 59.65              \\
        \hyperlink{https://huggingface.co/Qwen/Qwen-14B}{Qwen/Qwen-14B}                                                                                     & 55.95 & 90.73  & 63.16       & 22.61      & 38.32 & 64.91              \\
        \hyperlink{https://huggingface.co/Qwen/Qwen-14B-Chat}{Qwen/Qwen-14B-Chat}                                                                           & 54.71 & 91.56  & 65.79       & 22.36      & 32.42 & 61.40              \\
        \hyperlink{https://huggingface.co/rinna/nekomata-14b-instruction}{rinna/nekomata-14b-instruction}                                                   & 54.43 & 91.27  & 63.16       & 24.12      & 37.47 & 56.14              \\
        \hyperlink{https://huggingface.co/stabilityai/japanese-stablelm-base-beta-70b}{stabilityai/japanese-stablelm-base-beta-70b}                         & 53.07 & 90.87  & 60.53       & 22.36      & 33.68 & 57.89              \\
        \hyperlink{https://huggingface.co/stabilityai/japanese-stablelm-instruct-beta-70b}{stabilityai/japanese-stablelm-instruct-beta-70b}                 & 52.77 & 91.85  & 60.53       & 22.86      & 36.00 & 52.63              \\
        \hyperlink{https://huggingface.co/tokyotech-llm/Swallow-13b-instruct-hf}{tokyotech-llm/Swallow-13b-instruct-hf}                                     & 52.32 & 87.79  & 60.53       & 19.60      & 35.79 & 57.89              \\
        openai/gpt-35-turbo                                                                                                                                 & 50.27 & 89.98  & 52.63       & 18.09      & 29.26 & 61.40              \\
        \hyperlink{https://huggingface.co/meta-llama/Llama-2-70b-hf}{meta-llama/Llama-2-70b-hf}                                                             & 50.21 & 89.37  & 57.89       & 20.85      & 30.32 & 52.63              \\
        \hyperlink{https://huggingface.co/lightblue/qarasu-14B-chat-plus-unleashed}{lightblue/qarasu-14B-chat-plus-unleashed}                               & 50.04 & 89.69  & 57.89       & 20.35      & 31.37 & 50.88              \\
        \hyperlink{https://huggingface.co/rinna/nekomata-7b-instruction}{rinna/nekomata-7b-instruction}                                                     & 49.90 & 90.34  & 47.37       & 22.61      & 27.79 & 61.40              \\
        \hyperlink{https://huggingface.co/Qwen/Qwen-7B-Chat}{Qwen/Qwen-7B-Chat}                                                                             & 49.86 & 86.38  & 50.00       & 20.85      & 32.42 & 59.65              \\
        \hyperlink{https://huggingface.co/meta-llama/Llama-2-70b-chat-hf}{meta-llama/Llama-2-70b-chat-hf}                                                   & 49.53 & 90.29  & 52.63       & 18.84      & 28.00 & 57.89              \\
        \hyperlink{https://huggingface.co/Qwen/Qwen-7B}{Qwen/Qwen-7B}                                                                                       & 48.67 & 85.11  & 57.89       & 19.35      & 30.11 & 50.88              \\
        \hyperlink{https://huggingface.co/elyza/ELYZA-japanese-Llama-2-13b}{elyza/ELYZA-japanese-Llama-2-13b}                                               & 48.37 & 88.37  & 47.37       & 19.35      & 28.84 & 57.89              \\
        \hyperlink{https://huggingface.co/tokyotech-llm/Swallow-13b-hf}{tokyotech-llm/Swallow-13b-hf}                                                       & 48.31 & 87.59  & 52.63       & 19.60      & 32.63 & 49.12              \\
        \hyperlink{https://huggingface.co/Xwin-LM/Xwin-LM-13B-V0.2}{Xwin-LM/Xwin-LM-13B-V0.2}                                                               & 47.53 & 88.11  & 52.63       & 22.11      & 25.68 & 49.12              \\
        \hyperlink{https://huggingface.co/rinna/nekomata-7b}{rinna/nekomata-7b}                                                                             & 47.12 & 79.18  & 42.11       & 21.61      & 33.05 & 59.65              \\
        \hyperlink{https://huggingface.co/meta-llama/Llama-2-13b-chat-hf}{meta-llama/Llama-2-13b-chat-hf}                                                   & 46.98 & 87.95  & 52.63       & 19.60      & 27.37 & 47.37              \\
        \hyperlink{https://huggingface.co/elyza/ELYZA-japanese-Llama-2-7b-fast}{elyza/ELYZA-japanese-Llama-2-7b-fast}                                       & 46.04 & 82.52  & 44.74       & 17.84      & 30.74 & 54.39              \\
        \hyperlink{https://huggingface.co/elyza/ELYZA-japanese-Llama-2-13b-fast}{elyza/ELYZA-japanese-Llama-2-13b-fast}                                     & 45.70 & 86.37  & 39.47       & 20.60      & 31.16 & 50.88              \\
        \hyperlink{https://huggingface.co/lmsys/vicuna-13b-v1.5-16k}{lmsys/vicuna-13b-v1.5-16k}                                                             & 45.57 & 85.81  & 52.63       & 19.10      & 28.21 & 42.11              \\
        \hyperlink{https://huggingface.co/mosaicml/mpt-30b-instruct}{mosaicml/mpt-30b-instruct}                                                             & 45.18 & 83.27  & 42.11       & 21.36      & 26.53 & 52.63              \\
        \hyperlink{https://huggingface.co/meta-llama/Llama-2-7b-chat-hf}{meta-llama/Llama-2-7b-chat-hf}                                                     & 44.86 & 83.70  & 39.47       & 20.35      & 29.89 & 50.88              \\
        \hyperlink{https://huggingface.co/llm-jp/llm-jp-13b-instruct-full-jaster-v1.0}{llm-jp/llm-jp-13b-instruct-full-jaster-v1.0}                         & 44.66 & 85.91  & 39.47       & 20.10      & 26.95 & 50.88              \\
        \hyperlink{https://huggingface.co/elyza/ELYZA-japanese-Llama-2-13b-instruct}{elyza/ELYZA-japanese-Llama-2-13b-instruct}                             & 44.27 & 89.40  & 44.74       & 18.59      & 26.53 & 42.11              \\
        \hyperlink{https://huggingface.co/meta-llama/Llama-2-13b-hf}{meta-llama/Llama-2-13b-hf}                                                             & 44.19 & 82.04  & 36.84       & 20.85      & 30.32 & 50.88              \\
        \hyperlink{https://huggingface.co/rinna/youri-7b-instruction}{rinna/youri-7b-instruction}                                                           & 43.84 & 86.88  & 34.21       & 21.61      & 27.37 & 49.12              \\
        \hyperlink{https://huggingface.co/llm-jp/llm-jp-13b-instruct-full-dolly-oasst-v1.0}{llm-jp/llm-jp-13b-instruct-full-dolly-oasst-v1.0}               & 43.76 & 83.23  & 39.47       & 19.60      & 27.37 & 49.12              \\
        \hyperlink{https://huggingface.co/rinna/youri-7b-chat}{rinna/youri-7b-chat}                                                                         & 43.67 & 86.67  & 36.84       & 19.60      & 26.11 & 49.12              \\
        \hyperlink{https://huggingface.co/cyberagent/calm2-7b-chat}{cyberagent/calm2-7b-chat}                                                               & 43.67 & 81.09  & 36.84       & 18.09      & 29.68 & 52.63              \\
        \hyperlink{https://huggingface.co/llm-jp/llm-jp-13b-instruct-full-jaster-dolly-oasst-v1.0}{llm-jp/llm-jp-13b-instruct-full-jaster-dolly-oasst-v1.0} & 43.60 & 86.83  & 39.47       & 18.59      & 24.00 & 49.12              \\
        \hyperlink{https://huggingface.co/elyza/ELYZA-japanese-Llama-2-13b-fast-instruct}{elyza/ELYZA-japanese-Llama-2-13b-fast-instruct}                   & 43.59 & 87.27  & 42.11       & 18.59      & 26.11 & 43.86              \\
        \hyperlink{https://huggingface.co/lmsys/vicuna-33b-v1.3}{lmsys/vicuna-33b-v1.3}                                                                     & 43.44 & 87.81  & 34.21       & 19.60      & 28.21 & 47.37              \\
        \hyperlink{https://huggingface.co/lmsys/vicuna-7b-v1.5-16k}{lmsys/vicuna-7b-v1.5-16k}                                                               & 43.21 & 84.78  & 39.47       & 19.60      & 24.84 & 47.37              \\
        \hyperlink{https://huggingface.co/mosaicml/mpt-30b-chat}{mosaicml/mpt-30b-chat}                                                                     & 43.10 & 86.40  & 39.47       & 21.36      & 24.42 & 43.86              \\
        \hyperlink{https://huggingface.co/elyza/ELYZA-japanese-Llama-2-7b}{elyza/ELYZA-japanese-Llama-2-7b}                                                 & 42.99 & 83.48  & 42.11       & 19.60      & 25.89 & 43.86              \\
        \hyperlink{https://huggingface.co/tokyotech-llm/Swallow-7b-hf}{tokyotech-llm/Swallow-7b-hf}                                                         & 42.91 & 72.27  & 39.47       & 19.60      & 28.84 & 54.39              \\
        \hyperlink{https://huggingface.co/pfnet/plamo-13b}{pfnet/plamo-13b}                                                                                 & 42.87 & 76.97  & 39.47       & 21.61      & 27.16 & 49.12              \\
        \hyperlink{https://huggingface.co/mosaicml/mpt-30b}{mosaicml/mpt-30b}                                                                               & 42.80 & 83.44  & 36.84       & 19.60      & 26.74 & 47.37              \\
        \hyperlink{https://huggingface.co/stabilityai/japanese-stablelm-base-alpha-7b}{stabilityai/japanese-stablelm-base-alpha-7b}                         & 42.73 & 78.74  & 34.21       & 19.10      & 30.74 & 50.88              \\
        \hyperlink{https://huggingface.co/Xwin-LM/Xwin-LM-7B-V0.2}{Xwin-LM/Xwin-LM-7B-V0.2}                                                                 & 42.73 & 82.79  & 42.11       & 19.85      & 25.05 & 43.86              \\
        \hyperlink{https://huggingface.co/llm-jp/llm-jp-13b-v1.0}{llm-jp/llm-jp-13b-v1.0}                                                                   & 42.39 & 81.24  & 39.47       & 19.10      & 26.53 & 45.61              \\
        \hyperlink{https://huggingface.co/cyberagent/calm2-7b}{cyberagent/calm2-7b}                                                                         & 41.96 & 80.02  & 42.11       & 17.84      & 24.21 & 45.61              \\
        \hyperlink{https://huggingface.co/rinna/japanese-gpt-neox-3.6b-instruction-ppo}{rinna/japanese-gpt-neox-3.6b-instruction-ppo}                       & 41.89 & 74.71  & 44.74       & 20.60      & 23.79 & 45.61              \\
        \hyperlink{https://huggingface.co/rinna/youri-7b}{rinna/youri-7b}                                                                                   & 41.84 & 73.60  & 34.21       & 19.10      & 29.68 & 52.63              \\
        \hyperlink{https://huggingface.co/elyza/ELYZA-japanese-Llama-2-7b-fast-instruct}{elyza/ELYZA-japanese-Llama-2-7b-fast-instruct}                     & 41.59 & 82.53  & 39.47       & 20.10      & 25.47 & 40.35              \\
        \hyperlink{https://huggingface.co/stabilityai/japanese-stablelm-instruct-alpha-7b}{stabilityai/japanese-stablelm-instruct-alpha-7b}                 & 41.43 & 78.94  & 34.21       & 19.35      & 23.79 & 50.88              \\
        \hyperlink{https://huggingface.co/tokyotech-llm/Swallow-7b-instruct-hf}{tokyotech-llm/Swallow-7b-instruct-hf}                                       & 41.36 & 83.61  & 31.58       & 18.09      & 24.42 & 49.12              \\
        \hyperlink{https://huggingface.co/stabilityai/japanese-stablelm-instruct-alpha-7b-v2}{stabilityai/japanese-stablelm-instruct-alpha-7b-v2}           & 41.36 & 78.62  & 34.21       & 19.10      & 24.00 & 50.88              \\
        \hyperlink{https://huggingface.co/pfnet/plamo-13b-instruct}{pfnet/plamo-13b-instruct}                                                               & 41.13 & 77.33  & 39.47       & 21.11      & 27.37 & 40.35              \\
        \hyperlink{https://huggingface.co/rinna/japanese-gpt-neox-3.6b-instruction-sft-v2}{rinna/japanese-gpt-neox-3.6b-instruction-sft-v2}                 & 41.03 & 75.36  & 39.47       & 19.10      & 27.37 & 43.86              \\
        \hyperlink{https://huggingface.co/meta-llama/Llama-2-7b-hf}{meta-llama/Llama-2-7b-hf}                                                               & 40.99 & 77.41  & 39.47       & 18.59      & 27.37 & 42.11              \\
        \hyperlink{https://huggingface.co/rinna/bilingual-gpt-neox-4b-instruction-ppo}{rinna/bilingual-gpt-neox-4b-instruction-ppo}                         & 40.71 & 78.38  & 31.58       & 20.60      & 27.37 & 45.61              \\
        \hyperlink{https://huggingface.co/rinna/bilingual-gpt-neox-4b-instruction-sft}{rinna/bilingual-gpt-neox-4b-instruction-sft}                         & 40.31 & 78.23  & 34.21       & 19.35      & 25.89 & 43.86              \\
        \hyperlink{https://huggingface.co/llm-jp/llm-jp-1.3b-v1.0}{llm-jp/llm-jp-1.3b-v1.0}                                                                 & 39.70 & 75.48  & 36.84       & 19.85      & 24.21 & 42.11              \\
        At Random                                                                                                                                           & 30.68 & 49.15  & 25.00       & 16.98      & 25.00 & 37.28              \\
        \hline
    \end{tabular}
\end{table*}

\section{Experiments: Benchmark Calculation for LLMs}
We measured the benchmarks for various models using the benchmarks described in the previous section.

Given the significant impact of prompts on performance, we prepared prompts for each task in addition to the prompts presented in the previous section.
These prompts were similar to those employed in previous Japanese-specific benchmark studies \cite{jlm_eval}.
Preliminary experiments with 0--4 shots were conducted using these prompts, and the best-performing prompts and numbers of shots were employed for the final experiment.
Although this procedure may seem to be a type of in-sample training, in practice, we believe that such an evaluation procedure would provide a fair comparison.
This is because the number of prompts was limited, and it was easy for a human to train the model to select the most appropriate prompts.

However, for the models provided by Open AI through its API, we decided to use only one standard prompt and only 0-shots for the number of shots because of the cost.
The Open AI API was used with Azure; if a content filter was applied and no answer was obtained, it was determined to be incorrect.

To answer the multiple-choice questions, the likelihoods of the choices in the context were calculated and the choice with the highest likelihood was employed as the output.
For GPT3.5 and GPT-4 series, the outputs with the temperature parameter set to 0 were obtained via API, and the choice that appeared earliest in the outputs was used as the output.

The results are summarized in Table \ref{tab:benchmark}.

\section{Discussion}
According to the results, the GPT-4 series exhibited a significantly high performance.
Although the number of parameters in GPT-4 has not been determined, it is estimated to be more than 500 billion.
Compared with other models, which have approximately 70 billion or fewer parameters, the number of parameters in GPT-4 is significantly larger, at least a few times.
Considering that Qwen-72B exhibited the second-best results, the effect of the number of parameters in the models was important for achieving the highest results.

Compared to the existing Japanese leaderboard, Nejumi\footnote{\url{https://wandb.ai/wandb-japan/llm-leaderboard/reports/Nejumi-LLM-Neo--Vmlldzo2MTkyMTU0}}, our benchmark results for Japanese financial tasks almost correspond to the general Japanese task performance, but an exception exists.
Nekomata-14b exhibits a high performance in financial tasks, which differs from that of the Nejumi leaderboard.
Nekomata-14b is a tuned model of Qwen-14b that has not yet been evaluated on the Nejumi leaderboard.
Moreover, the training corpora for the Qwen series were not revealed, but corpora of professional fields were included according to the official website.
Therefore, the corpora used in the training of Qwen may include financial-related texts in their pre-training, and the performance of nekomata-14b is owing to this.
However, models other than the nekomata, Qwen, and GPT series are already known to not include financial-related texts in their pre-training.

In the middle score of the benchmarks, around the model exhibiting an overall score of 35--40, no significant differences in their performances or the effect of the number of parameters in the models were present.
We believe that this is also related to the corpora used in the training of the models.
Currently, several LLMs do not learn financial documents.
Therefore, in the future, the impact of financial texts on training should be evaluated, and developing models trained with financial documents is also important.

From the overall summary of the results, the benchmarks that we constructed exhibited considerable variation in difficulty from task to task, and it is possible that we were making an effective assessment.
With respect to Chabsa, the highest-performing models approached the theoretical upper limit.
For the design of this task, we believe that 95 is a realistic upper limit that can be achieved and is almost at this limit.
However, room for further improvement in other tasks still exists, specifically regarding the performance of cpa\_audit.
A previous study \citeplanguageresource{Masuda2023} reported that a combination of GPT-4 and retrieval-augmented generation is necessary to achieve a passing level of performance.
The model's performance in solving the cpa\_audit task without any external information sources can still be improved.

\begin{figure*}[tb]
    \centering
    \begin{minipage}[t]{0.3\linewidth}
        \centering
        \includegraphics[width=\linewidth]{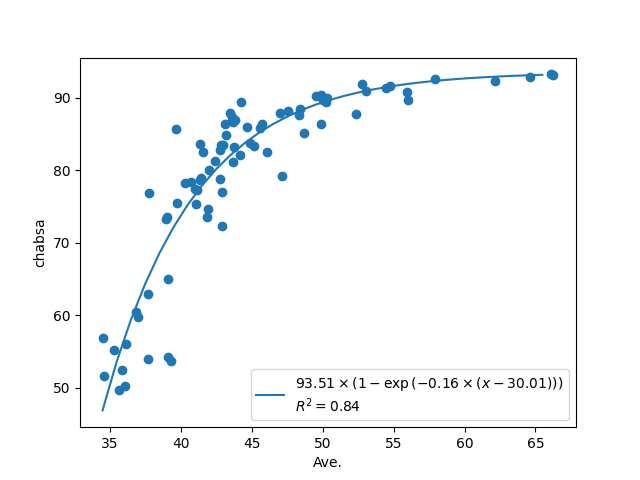}
        \caption{Relationship between Benchmark and chabsa scores}
        \label{fig:fig1}
    \end{minipage}
    ~~~
    \begin{minipage}[t]{0.3\linewidth}
        \centering
        \includegraphics[width=\linewidth]{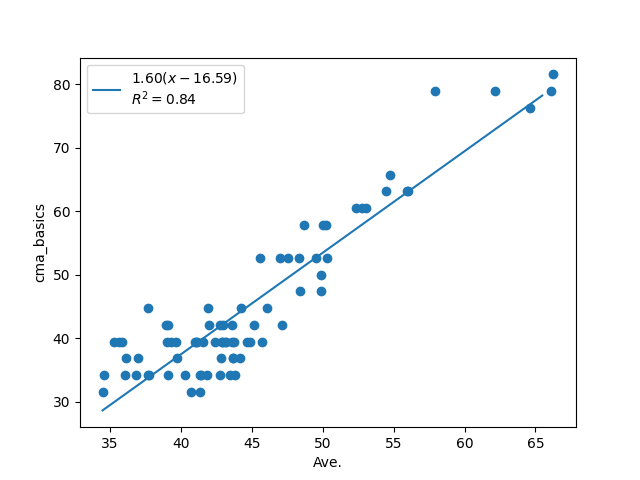}
        \caption{Relationship between Benchmark and cma\_basics scores}
        \label{fig:fig2}
    \end{minipage}
    ~~~
    \begin{minipage}[t]{0.3\linewidth}
        \centering
        \includegraphics[width=\linewidth]{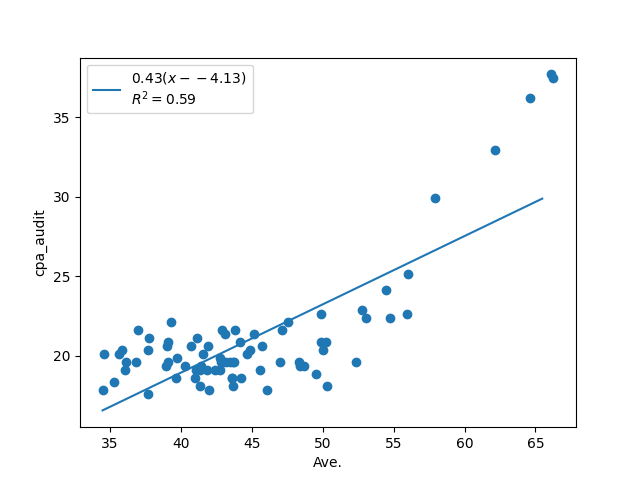}
        \caption{Relationship between Benchmark and cpa\_audit scores}
        \label{fig:fig3}
    \end{minipage}
    \begin{minipage}[t]{0.33\linewidth}
        \centering
        \includegraphics[width=\linewidth]{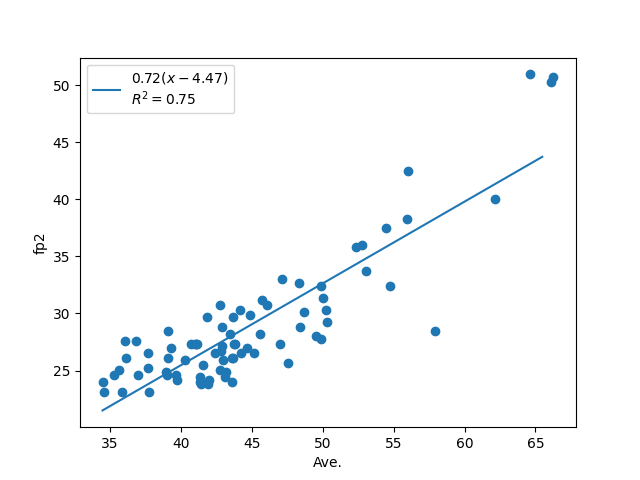}
        \caption{Relationship between Benchmark and fp2 scores}
        \label{fig:fig4}
    \end{minipage}
    ~~~
    \begin{minipage}[t]{0.4\linewidth}
        \centering
        \includegraphics[width=0.83\linewidth]{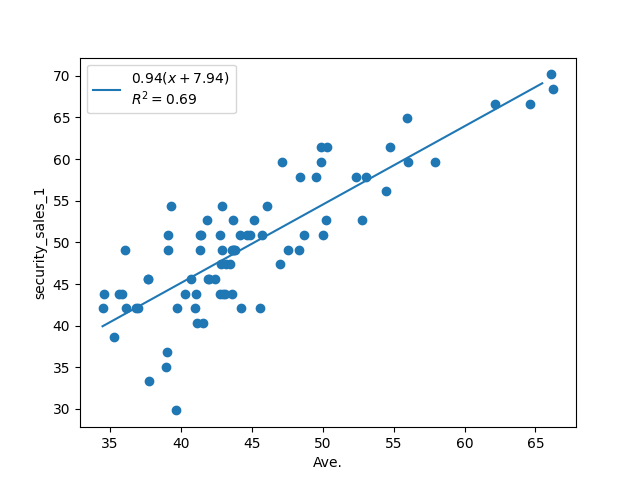}
        \caption{Relationship between Benchmark and security\_sales\_1 scores}
        \label{fig:fig5}
    \end{minipage}
\end{figure*}

To investigate the effectivity of our benchmark, we analyzed the results, and the plots shown in Figures \ref{fig:fig1} -- \ref{fig:fig5} were created.
The relationships between the overall benchmark score and the individual scores for each task are plotted in Figures \ref{fig:fig1} -- \ref{fig:fig5}.
Because 1/5 of the mean score is obtained from each task, a certain degree of correlation can be observed.
In Figure \ref{fig:fig1}, the scatter plot appears to be similar to that of $1-\exp{(x)}$; therefore, fitting was performed using that function.
This implies that the task tended to be easy and saturated for higher-performing models.
The fitting function was found to fit well.

According to the plots, each task has its own difficulties.
Chabsa is a relatively easy task and a good indicator that the difference in scores widens in lower-performing tiers.
In addition, for cma\_basics and security\_sales\_1, there is little difference in the scores of the lower-performing tiers, but the difference in the scores of the mid-performing tiers is increasing.
In contrast, for the other indicators, that is, cpa\_audit and fp2, observing differences in performance for both the lower and middle-performing tiers is difficult, and only some of the models exhibit overwhelmingly high performance.
Because of the inclusion of these tasks with varying difficulty levels, our constructed benchmarks seem to be suitable for evaluating the Japanese financial performance of LLMs.

In future studies, we need to add more tasks, introduce more reasonable prompt-tuning methods, and determine whether a finance-specific language model can perform well.

\section{Conclusion}
In this study, we constructed a new LLM benchmark specialized for Japanese financial tasks and measured the actual benchmarks for various models.
The results demonstrated that the GPT-4 series exhibited overwhelming performance.
In contrast, we were also able to confirm the usefulness of our benchmark.
We confirmed that our benchmark could differentiate the benchmark scores among models in all performance ranges by combining tasks with different difficulties.
Future studies should also include more tasks for benchmarking to ensure a more accurate performance evaluation of LLMs.

\section*{Declarations}
The author is affiliated with Preferred Networks, Inc., the developer of \hyperlink{https://huggingface.co/pfnet/plamo-13b}{pfnet/plamo-13b}, \hyperlink{https://huggingface.co/pfnet/plamo-13b-instruct}{pfnet/plamo-13b-instruct}, and \hyperlink{https://huggingface.co/pfnet/plamo-13b-instruct-nc}{pfnet/plamo-13b-instruct-nc}.
However, in the experiments conducted in this study, all codes were made publicly available for transparency and fair evaluation with other models.

\nocite{*}
\section{Bibliographical References}\label{sec:reference}

\bibliographystyle{lrec-coling2024-natbib}
\bibliography{cite}

\begin{thebibliography}{2}
\expandafter\ifx\csname natexlab\endcsname\relax\def\natexlab#1{#1}\fi

\bibitem[{Kubo et~al.(2018)Kubo, Nakayama, and Kamura}]{chabsa}
Kubo, Takahiro and Nakayama, Hiroki and Kamura, Junya. 2018.
\newblock \emph{{chABSA: Aspect Based Sentiment Analysis dataset in Japanese}}.
\newblock PID \href{https://github.com/chakki-works/chABSA-dataset}{https://github.com/chakki-works/chABSA-dataset}.

\bibitem[{Masuda et~al.(2023)Masuda, Nakagawa, and Hoshino}]{Masuda2023}
Tatsuki Masuda, Kei Nakagawa, and Takahiro Hoshino. 2023.
\newblock Can chatgpt pass the jcpa exam?: Challenge for the short-answer method test on auditing.
\newblock In \emph{The 31st meeting of Special Interest Group on Financial Informatics of Japanese Society for Artificial Intelligence}, pages 81--88.

\end{thebibliography}


\begin{thebibliography}{40}
\expandafter\ifx\csname natexlab\endcsname\relax\def\natexlab#1{#1}\fi

\bibitem[{Anil et~al.(2023)Anil, Dai et~al.}]{Anil2023}
Rohan Anil, Andrew~M. Dai, et~al. 2023.
\newblock {PaLM 2 Technical Report}.
\newblock \emph{arXiv}.
\newblock \url{https://arxiv.org/abs/2305.10403v3}.

\bibitem[{Beltagy et~al.(2019)Beltagy, Lo, and Cohan}]{beltagy2019scibert}
Iz~Beltagy, Kyle Lo, and Arman Cohan. 2019.
\newblock Scibert: A pretrained language model for scientific text.
\newblock In \emph{Proceedings of the 2019 Conference on Empirical Methods in Natural Language Processing and the 9th International Joint Conference on Natural Language Processing (EMNLP-IJCNLP)}, pages 3615--3620.

\bibitem[{Brown et~al.(2020)Brown, Mann et~al.}]{GPT-3}
Tom Brown, Benjamin Mann, et~al. 2020.
\newblock {Language Models are Few-Shot Learners}.
\newblock \emph{Advances in Neural Information Processing Systems}, 33:1877--1901.

\bibitem[{Choi et~al.(2023)Choi, Hickman, Monahan, and Schwarcz}]{Choi2023}
Jonathan~H. Choi, Kristin~E. Hickman, Amy Monahan, and Daniel~B. Schwarcz. 2023.
\newblock \href {https://doi.org/10.2139/SSRN.4335905} {{ChatGPT Goes to Law School}}.
\newblock \emph{SSRN Electronic Journal}.
\newblock \url{https://papers.ssrn.com/abstract=4335905}.

\bibitem[{Chowdhery et~al.(2022)Chowdhery, Narang et~al.}]{Chowdhery2022}
Aakanksha Chowdhery, Sharan Narang, et~al. 2022.
\newblock {PaLM: Scaling Language Modeling with Pathways}.
\newblock \emph{arXiv}.
\newblock \url{https://arxiv.org/abs/2204.02311v5}.

\bibitem[{Databricks(2023)}]{dolly}
Databricks. 2023.
\newblock Dolly.
\newblock \url{https://github.com/databrickslabs/dolly}.

\bibitem[{Devlin et~al.(2019)Devlin, Chang, Lee, and Toutanova}]{Devlin2018}
Jacob Devlin, Ming-Wei Chang, Kenton Lee, and Kristina Toutanova. 2019.
\newblock \href {https://doi.org/10.18653/v1/N19-1423} {{BERT: Pre-training of Deep Bidirectional Transformers for Language Understanding}}.
\newblock In \emph{Proceedings of the 2019 Conference of the North {A}merican Chapter of the Association for Computational Linguistics}, pages 4171--4186. Association for Computational Linguistics.

\bibitem[{Eulerich et~al.(2023)Eulerich, Sanatizadeh, Vakilzadeh, and Wood}]{Eulerich2023}
Marc Eulerich, Aida Sanatizadeh, Hamid Vakilzadeh, and David~A. Wood. 2023.
\newblock \href {https://doi.org/10.2139/SSRN.4452175} {{Is it All Hype? ChatGPT’s Performance and Disruptive Potential in the Accounting and Auditing Industries}}.
\newblock \emph{SSRN Electronic Journal}.
\newblock \url{https://papers.ssrn.com/abstract=4452175}.

\bibitem[{Gao et~al.(2021)Gao, Tow, Biderman, Black, DiPofi et~al.}]{eval-harness}
Leo Gao, Jonathan Tow, Stella Biderman, Sid Black, Anthony DiPofi, et~al. 2021.
\newblock \href {https://doi.org/10.5281/zenodo.5371628} {A framework for few-shot language model evaluation}.
\newblock \url{https://github.com/EleutherAI/lm-evaluation-harness}.

\bibitem[{Google(2023)}]{bard}
Google. 2023.
\newblock Bard.
\newblock \url{https://bard.google.com/}.

\bibitem[{Gururangan et~al.(2020)Gururangan, Marasovi{\'c}, Swayamdipta, Lo, Beltagy, Downey, and Smith}]{gururangan2020don}
Suchin Gururangan, Ana Marasovi{\'c}, Swabha Swayamdipta, Kyle Lo, Iz~Beltagy, Doug Downey, and Noah~A Smith. 2020.
\newblock Don't stop pretraining: Adapt language models to domains and tasks.
\newblock \emph{arXiv preprint arXiv:2004.10964}.

\bibitem[{HIRANO et~al.(2019)HIRANO, SAKAJI, KIMURA, IZUMI, MATSUSHIMA, NAGAO, and KATO}]{Hirano2019-information}
Masanori HIRANO, Hiroki SAKAJI, Shoko KIMURA, Kiyoshi IZUMI, Hiroyasu MATSUSHIMA, Shintaro NAGAO, and Atsuo KATO. 2019.
\newblock \href {https://doi.org/10.3390/info10030102} {{Related Stocks Selection with Data Collaboration Using Text Mining}}.

\bibitem[{HIRANO et~al.(2023)HIRANO, SUZUKI, and SAKAJI}]{Hirano2023-nbis}
Masanori HIRANO, Masahiro SUZUKI, and Hiroki SAKAJI. 2023.
\newblock \href {https://doi.org/10.1007/978-3-031-40978-3_47} {{llm-japanese-dataset v0: Construction of Japanese Chat Dataset for Large Language Models and its Methodology}}.
\newblock In \emph{The 26th International Conference on Network-Based Information Systems}, pages 442--454.

\bibitem[{Howard and Ruder(2018)}]{howard2018ulmfit}
Jeremy Howard and Sebastian Ruder. 2018.
\newblock \href {https://doi.org/10.18653/v1/P18-1031} {Universal language model fine-tuning for text classification}.
\newblock In \emph{Proceedings of the 56th Annual Meeting of the Association for Computational Linguistics (Volume 1: Long Papers)}, pages 328--339. Association for Computational Linguistics.

\bibitem[{Iu and Wong(2023)}]{Iu2023}
Kwan~Yuen Iu and Vanessa Man-Yi Wong. 2023.
\newblock \href {https://doi.org/10.2139/SSRN.4339839} {{ChatGPT by OpenAI: The End of Litigation Lawyers?}}
\newblock \emph{SSRN Electronic Journal}.
\newblock \url{https://papers.ssrn.com/abstract=4339839}.

\bibitem[{LLM-jp(2024)}]{llm-jp-eval}
LLM-jp. 2024.
\newblock \href {https://github.com/llm-jp/llm-jp-eval} {{llm-jp-eval}}.

\bibitem[{Mikolov et~al.(2013)Mikolov, Chen, Corrado, and Dean}]{Mikolov2013a}
Tomas Mikolov, Kai Chen, Greg Corrado, and Jeffrey Dean. 2013.
\newblock \href {http://papers.nips.cc/paper/5021-distributed-representations-of-words-and-phrases-and-their-compositionality} {{Distributed Representations of Words and Phrases and their Compositionality}}.
\newblock In \emph{Advances in Neural Information Processing Systems (NeurIPS)}, volume~26, pages 3111--3119.

\bibitem[{Nori et~al.(2023)Nori, King, Mckinney, Carignan, and Horvitz}]{Nori2023}
Harsha Nori, Nicholas King, Scott~Mayer Mckinney, Dean Carignan, and Eric Horvitz. 2023.
\newblock {Capabilities of GPT-4 on Medical Challenge Problems}.
\newblock \emph{arXiv}.
\newblock \url{https://arxiv.org/abs/2303.13375v2}.

\bibitem[{OpenAI(2023{\natexlab{a}})}]{chatgpt}
OpenAI. 2023{\natexlab{a}}.
\newblock {ChatGPT}.
\newblock \url{https://openai.com/blog/chatgpt/}.

\bibitem[{OpenAI(2023{\natexlab{b}})}]{GPT4}
OpenAI. 2023{\natexlab{b}}.
\newblock \href {http://arxiv.org/abs/2303.08774} {{GPT-4 Technical Report}}.

\bibitem[{Peters et~al.(2018)Peters, Neumann, Iyyer, Gardner, Clark, Lee, and Zettlemoyer}]{peters2018elmo}
Matthew~E. Peters, Mark Neumann, Mohit Iyyer, Matt Gardner, Christopher Clark, Kenton Lee, and Luke Zettlemoyer. 2018.
\newblock \href {https://doi.org/10.18653/v1/N18-1202} {Deep contextualized word representations}.
\newblock In \emph{Proceedings of the 2018 Conference of the North {A}merican Chapter of the Association for Computational Linguistics: Human Language Technologies}, volume~1, pages 2227--2237. Association for Computational Linguistics.

\bibitem[{Radford et~al.(2018)Radford, Narasimhan, Salimans, and Sutskever}]{GPT-1}
Alec Radford, Karthik Narasimhan, Tim Salimans, and Ilya Sutskever. 2018.
\newblock {Improving Language Understanding by Generative Pre-Training}.
\newblock \url{https://cdn.openai.com/research-covers/language-unsupervised/language_understanding_paper.pdf}.

\bibitem[{Radford et~al.(2019)Radford, Wu, Child, Luan, Amodei, and Sutskever}]{GPT-2}
Alec Radford, Jeff Wu, Rewon Child, David Luan, Dario Amodei, and Ilya Sutskever. 2019.
\newblock {Language Models are Unsupervised Multitask Learners}.
\newblock \url{https://cdn.openai.com/better-language-models/language_models_are_unsupervised_multitask_learners.pdf}.

\bibitem[{Rasmy et~al.(2021)Rasmy, Xiang, Xie, Tao, and Zhi}]{rasmy2021med}
Laila Rasmy, Yang Xiang, Ziqian Xie, Cui Tao, and Degui Zhi. 2021.
\newblock Med-bert: pretrained contextualized embeddings on large-scale structured electronic health records for disease prediction.
\newblock \emph{NPJ digital medicine}, 4(1):86.

\bibitem[{Scao et~al.(2022)Scao, Fan, Akiki, Pavlick, Ili{\'c}, Hesslow, Castagn{\'e}, Luccioni, Yvon, Gall{\'e} et~al.}]{scao2022bloom}
Teven~Le Scao, Angela Fan, Christopher Akiki, Ellie Pavlick, Suzana Ili{\'c}, Daniel Hesslow, Roman Castagn{\'e}, Alexandra~Sasha Luccioni, Fran{\c{c}}ois Yvon, Matthias Gall{\'e}, et~al. 2022.
\newblock {BLOOM: A 176B-Parameter Open-Access Multilingual Language Model}.
\newblock \emph{arXiv}.
\newblock \url{https://arxiv.org/abs/2211.05100}.

\bibitem[{Schuster and Paliwal(1997)}]{schuster1997bilstm}
M.~Schuster and K.K. Paliwal. 1997.
\newblock \href {https://doi.org/10.1109/78.650093} {Bidirectional recurrent neural networks}.
\newblock \emph{IEEE Transactions on Signal Processing}, 45(11):2673--2681.

\bibitem[{StabilityAI(2023)}]{jlm_eval}
StabilityAI. 2023.
\newblock {JP Language Model Evaluation Harness}.
\newblock \url{https://github.com/Stability-AI/lm-evaluation-harness/tree/jp-stable}.

\bibitem[{Sukeda et~al.(2023)Sukeda, Suzuki, Sakaji, and Kodera}]{sukeda2023jmedlora}
Issey Sukeda, Masahiro Suzuki, Hiroki Sakaji, and Satoshi Kodera. 2023.
\newblock {JMedLoRA: Medical Domain Adaptation on Japanese Large Language Models using Instruction-tuning}.
\newblock \emph{arXiv}.
\newblock \url{https://arxiv.org/abs/2310.10083}.

\bibitem[{Suzuki et~al.(2023)Suzuki, Hirano, and Sakaji}]{suzuki2023base}
Masahiro Suzuki, Masanori Hirano, and Hiroki Sakaji. 2023.
\newblock {From Base to Conversational: Japanese Instruction Dataset and Tuning Large Language Models}.
\newblock \emph{arXiv}.
\newblock \url{https://arxiv.org/abs/2309.03412}.

\bibitem[{SUZUKI et~al.(2022)SUZUKI, SAKAJI, HIRANO, and IZUMI}]{Suzuki2022-sigfin28}
Masahiro SUZUKI, Hiroki SAKAJI, Masanori HIRANO, and Kiyoshi IZUMI. 2022.
\newblock \href {https://doi.org/10.11517/jsaisigtwo.2022.FIN-028_132} {{Construction and Validation of a Pre-Training and Additional Pre-Training Financial Language Model [in Japanese]}}.
\newblock In \emph{The 28th meeting of Special Interest Group on Financial Informatics of Japanese Society for Artificial Intelligence}, pages 132--137.

\bibitem[{SUZUKI et~al.(2023)SUZUKI, SAKAJI, HIRANO, and IZUMI}]{Suzuku2023-ipm}
Masahiro SUZUKI, Hiroki SAKAJI, Masanori HIRANO, and Kiyoshi IZUMI. 2023.
\newblock \href {https://doi.org/10.1016/j.ipm.2022.103194} {{Constructing and Analyzing Domain-Specific Language Model for Financial Text Mining}}.

\bibitem[{Team(2023)}]{gemini}
Gemini Team. 2023.
\newblock Gemini: a family of highly capable multimodal models.
\newblock \emph{arXiv preprint arXiv:2312.11805}.

\bibitem[{Touvron et~al.(2023{\natexlab{a}})Touvron, Lavril, Izacard, Martinet, Lachaux, Lacroix, Rozi{\`e}re, Goyal, Hambro, Azhar et~al.}]{touvron2023llama}
Hugo Touvron, Thibaut Lavril, Gautier Izacard, Xavier Martinet, Marie-Anne Lachaux, Timoth{\'e}e Lacroix, Baptiste Rozi{\`e}re, Naman Goyal, Eric Hambro, Faisal Azhar, et~al. 2023{\natexlab{a}}.
\newblock {LLaMA: Open and Efficient Foundation Language Models}.
\newblock \emph{arXiv}.
\newblock \url{https://arxiv.org/abs/2302.13971}.

\bibitem[{Touvron et~al.(2023{\natexlab{b}})Touvron, Martin et~al.}]{Touvron2023}
Hugo Touvron, Louis Martin, et~al. 2023{\natexlab{b}}.
\newblock {Llama 2: Open Foundation and Fine-Tuned Chat Models}.
\newblock \emph{arXiv}.
\newblock \url{https://arxiv.org/abs/2307.09288v2}.

\bibitem[{Vaswani et~al.(2017)Vaswani, Shazeer, Parmar, Uszkoreit, Jones, Gomez, Kaiser, and Polosukhin}]{Vaswani2017}
Ashish Vaswani, Noam Shazeer, Niki Parmar, Jakob Uszkoreit, Llion Jones, Aidan~N. Gomez, {\L}ukasz Kaiser, and Illia Polosukhin. 2017.
\newblock {Attention Is All You Need}.
\newblock In \emph{Advances in Neural Information Processing Systems}, volume~30, pages 5999--6009.

\bibitem[{Vicuna(2023)}]{vicuna}
Vicuna. 2023.
\newblock {Vicuna: An Open-Source Chatbot Impressing GPT-4 with 90\%* ChatGPT Quality}.
\newblock \url{https://vicuna.lmsys.org/}.

\bibitem[{William~Todt(2023)}]{Fin-LLAMA}
Pedram~Babaei William~Todt, Ramtin~Babaei. 2023.
\newblock {Fin-LLAMA: Efficient Finetuning of Quantized LLMs for Finance}.
\newblock \url{https://github.com/Bavest/fin-llama}.

\bibitem[{Wu et~al.(2023)Wu, Irsoy, Lu, Dabravolski, Dredze, Gehrmann, Kambadur, Rosenberg, and Mann}]{Wu2023}
Shijie Wu, Ozan~˙ Irsoy, Steven Lu, Vadim Dabravolski, Mark Dredze, Sebastian Gehrmann, Prabhanjan Kambadur, David Rosenberg, and Gideon Mann. 2023.
\newblock {BloombergGPT: A Large Language Model for Finance}.
\newblock \emph{arXiv}.
\newblock \url{https://arxiv.org/abs/2303.17564v2}.

\bibitem[{Yang et~al.(2023)Yang, Liu, and Wang}]{yang2023fingpt}
Hongyang Yang, Xiao-Yang Liu, and Christina~Dan Wang. 2023.
\newblock {FinGPT: Open-Source Financial Large Language Models}.
\newblock \emph{arXiv}.
\newblock \url{https://arxiv.org/abs/2306.06031}.

\bibitem[{Zhang et~al.(2023)Zhang, Yang, and Liu}]{zhang2023instruct}
Boyu Zhang, Hongyang Yang, and Xiao-Yang Liu. 2023.
\newblock {Instruct-FinGPT: Financial Sentiment Analysis by Instruction Tuning of General-Purpose Large Language Models}.
\newblock \emph{arXiv}.
\newblock \url{https://arxiv.org/abs/2306.12659}.

\end{thebibliography}

\section{Language Resource References}
\label{lr:ref}
\bibliographystylelanguageresource{lrec-coling2024-natbib}
\bibliographylanguageresource{languageresource}

\end{document}